\documentclass[ALICE,manyauthors]{cernphprep}

\usepackage[comma,square,numbers,sort&compress]{natbib}
\usepackage[hyperfootnotes=false]{hyperref}

\usepackage{multirow}
\usepackage{bm}
\usepackage{diagbox}
\usepackage{makecell}
\usepackage{color}
\usepackage{graphicx}
\usepackage[T1]{fontenc}

\newcommand{\PsiRP}{\ensuremath{\Psi_{\rm RP}}\xspace}
\newcommand{\psisp}{\Psi_{\rm SP}}
\newcommand{\Rep}{\ensuremath{R_{\rm SP}^{(1)}}\xspace}

\newcommand{\ph}{\ensuremath{P_{\rm H}}\xspace}
\newcommand{\pt}{p_{\rm T}}

\newcommand{\Qv}{{\rm \bf Q}}
\newcommand{\tpspec}{{\rm t,p}}
\newcommand{\Xpm}{\Qx^\tpspec}
\newcommand{\Ypm}{\Qy^\tpspec}
\newcommand{\Qtp}{\Qv^\tpspec}

\newcommand{\Qx}{Q_{\rm x}}
\newcommand{\Qy}{Q_{\rm y}}

\newcommand{ \la }{\langle}
\newcommand{ \ra }{\rangle}
\newcommand{ \bs }{{\bf s}}

\newcommand{ \bv }{{\bf v}}

\newcommand{\mean}[1]{\la #1 \ra}

\newcommand{ \grad }{{\bf \nabla}}

\newcommand{ \bomega }{{\boldsymbol{\omega}}}

\definecolor{dgreen}{cmyk}{1.,0.,1.,0.2} % dark green
\definecolor{orange}{cmyk}{0.,0.353,1.,0.} % orange

\newcommand{\pT}{\ensuremath{p_{\rm T}}\xspace}

\newcommand{\sNN}{\ensuremath{\sqrt{s_{\rm NN}}}\xspace}

\newcommand{\Minv}{\ensuremath{\mathrm{M}_\mathrm{inv}}\xspace}
\newcommand{\fBG}{\ensuremath{f_\mathrm{BG}(\Minv)}\xspace}

\begin{document}%

%%%%%%%%%%%%%%%  Title page %%%%%%%%%%%%%%%%%%%%%%%%
\begin{titlepage}
% CERN-EP-2019-173
\PHyear{2019}       % required, will be obtained from CERN
\PHnumber{173}      % required, will be obtained from CERN
\PHdate{8 August}  % required, will be obtained from CERN
%%%%%%%%%%%%%%%%%%%%%%%%%%%%%%%%%%%%%%%%%%%%%%%%%%%%

%%% Put your own title + short title here:
\title{Global polarization of $\Lambda$ and $\overline{\Lambda}$ hyperons in Pb--Pb collisions at the LHC}

\ShortTitle{Global polarization of hyperons in Pb--Pb collisions at the LHC}
% appears on right page headers

%%% Do not change the next lines
\Collaboration{ALICE Collaboration\thanks{See Appendix~\ref{app:collab} for the list of collaboration members}}
\ShortAuthor{ALICE Collaboration} 
% appears on left page headers, do not change

\begin{abstract}
The global polarization of the $\Lambda$ and $\overline\Lambda$ hyperons is measured for Pb--Pb collisions at $\sNN=2.76$ and 5.02~TeV recorded with the ALICE at the LHC.
The results are reported differentially as a function of collision centrality and hyperon's transverse momentum (\pT) for the range of  centrality 5--50\%, $0.5 < \pT<5$~GeV/$c$, and rapidity $|y|<0.5$.
The hyperon global polarization averaged for Pb--Pb collisions at $\sNN=2.76$ and 5.02~TeV  is found to be consistent with zero, $\mean{\ph}(\%)\approx -0.01\pm 0.05~\mbox{(stat.)}  \pm 0.03~\mbox{(syst.)}$ in the collision centrality range 15--50\%, where the largest signal is expected.
The results are compatible with expectations based on an extrapolation from measurements at lower collision energies at RHIC, hydrodynamical model calculations, and empirical estimates based on collision energy dependence of directed flow, all of which predict the global polarization values at LHC energies of the order of $0.01\%$.
\end{abstract}

\end{titlepage}
\setcounter{page}{2}

\section{Introduction}

The system created in a non-central nucleus--nucleus collision might retain a significant fraction of the large orbital angular momentum of the colliding nuclei.
Due to the spin-orbit coupling, particles produced in such a collision can become globally polarized~\cite{Liang:2004ph,Voloshin:2004ha,Liang:2004xn,Gao:2007bc}.
The global polarization is a phenomena when spins of all emitted final state particles in a non-central nucleus--nucleus collision are aligned along one preferential direction defined by the initial angular orbital momentum. 
Its measurement provides important information about the initial conditions and dynamics of the quark-gluon plasma (QGP), as well as the hadronization process~\cite{Becattini:2015ska,Li:2017slc,Xie:2017upb}.
The global polarization for a specific particle and corresponding anti-particle is expected to be very similar, or even identical, in a system with small or zero baryon chemical potential.

High-energy heavy-ion collisions are also characterized by ultra-strong magnetic fields~\cite{Skokov:2009qp,Tuchin:2013ie}, which on average are aligned with the direction of the angular momentum.
While the peak values of the magnetic fields, which reach up to $\rm 10^{18}$ Gauss, can be estimated rather accurately~\cite{Skokov:2009qp,Tuchin:2013ie}, the time evolution, which depends on the QGP electric conductivity, is practically unknown.
These fields can also contribute to the global polarization, but their action on particles and anti-particles is expected to be in an opposite direction.
Thus the measurement of the splitting between particle and anti-particle global polarizations provides very valuable information about the QGP properties.

The measurements of the global polarization \ph for spin one half $\Lambda$ ($\overline\Lambda$) strange hyperons are experimentally favourable because their spin direction can be reconstructed via their weak decay topology into proton (anti-proton) and charged pion.
Recently, the STAR Collaboration observed non-zero global polarization of $\Lambda$ and $\overline\Lambda$ hyperons in Au--Au collisions, first for the RHIC beam energy scan (BES) energies of $\sNN = 7-39$~GeV and, later, with more data, at $\sNN = 200$~GeV~\cite{STAR:2017ckg,Adam:2018ivw}.
The magnitude of the observed polarization varies from a few to a fraction of a percent.
While the \ph values for $\Lambda$ and $\overline\Lambda$ agree within the experimental uncertainties, they are systematically higher for $\overline\Lambda$ than for $\Lambda$.
Assuming that this difference originates due to the magnetic fields one estimates the field strength in units of elementary charge $e$ to be $eB\sim 0.01 m_\pi^2$~\cite{Becattini:2016gvu}.

The exact nature of the spin-orbit interaction leading to the global polarization is not known.
It is unclear at what stage of the system evolution (the QGP, hadronization, or the hadronic rescattering) the polarization is acquired, neither the corresponding relaxation times are known.
Most of the recent calculations of the global polarization assume complete thermal equilibrium and validity of the hydrodynamical description of the system~\cite{Karpenko:2016jyx,Becattini:2016gvu,Florkowski:2017ruc,Li:2017slc,Baznat:2017jfj}.
They relate the particle polarization to the system's thermal vorticity at the hadronization time.
In a non-relativistic limit, assuming complete thermal equilibrium, the polarization of the particles can be evaluated as $\bm{\zeta}=\mean{\bs}/s=(s+1)\bm{\omega}/(3T)$, where $s$ is the particle's spin, $T$ is the system temperature, $\bomega=(1/2)(\grad\times \bv)$ is a nonrelativistic vorticity, and $\bv$ is the local fluid velocity~\cite{Becattini:2016gvu}.

The global polarization is determined by the average vorticity component perpendicular to the collision reaction plane, which is spanned by the beam direction and the impact parameter vector.
The global polarization measurements with the produced particle provide important information on both the nature of the spin-orbit interaction and the profile of velocity fields of the expanding system.
Both, the magnitude and the direction of the vorticity, can strongly vary within the system~\cite{Voloshin:2017kqp}.
In particular, a significant component along the beam direction can be acquired due to the transverse anisotropic flow~\cite{Voloshin:2017kqp,Becattini:2017gcx}.

The vorticity of the system, especially its component along the system's orbital momentum, is directly related to the asymmetries in the initial velocity fields.
This links the vorticity with the directed flow $v_1$, that is also strongly dependent on those asymmetries. The $v_1$ is defined by the first Fourier moment $v_1 = \langle \cos (\varphi-\PsiRP)\rangle$ of the produced particle's azimuthal asymmetry relative to the collision reaction plane angle \PsiRP.
Hydrodynamic simulations show that the orbital angular momentum stored in the system and the directed flow of charged particles are almost directly proportional to each other~\cite{Becattini:2015ska}.
This allows for an empirical estimate of the collision energy dependence of the global polarization~\cite{Voloshin:2017kqp}.
The STAR results for the directed flow~\cite{Adamczyk:2014ipa,Shanmuganathan:2015qxb} and the hyperon global polarization~\cite{STAR:2017ckg,Adam:2018ivw} from the BES program show that the slopes of $v_1$ at midrapidity (${\rm d}v_1/{\rm d}\eta$) for charged hadrons (pions) and the hyperon polarization are indeed strongly correlated.
The charged-particle directed flow in Pb--Pb collisions at $\sNN=2.76$~TeV is about three times smaller~\cite{Abelev:2013cva} than at the top RHIC energy of 200 GeV.
This suggest that the global polarization at the LHC energies should be also about three times smaller than at RHIC (around $\sim 0.08$\%) and decreasing from $\sNN = 2.76$~TeV to 5.02~TeV  by about $\sim$30\%~\cite{Margutti:2017lup}.
Even smaller polarization values at the LHC are expected when the directed flow is seen as a combination of the two effects -- the tilt of the source in the longitudinal direction and the dipole flow originating from the asymmetry in the initial energy density distributions~\cite{Adamczyk:2017ird}.
Taking into account that only the contribution to the directed flow from the tilted source is related to the vorticity and that its contribution relative to the dipole flow decreases with the collision energy~\cite{Adamczyk:2017ird}, one arrives to an estimate for the global polarization at the LHC energies of the order of $\sim0.04$\%.

In this paper, the measurements of the global polarization of the $\Lambda$ and $\overline\Lambda$ hyperons in Pb--Pb collisions at $\sNN=2.76$ and 5.02~TeV recorded with ALICE at the LHC are reported.
The paper is organized as follows.
In Sec.~\ref{observable} the analysis details are presented, and the global polarization observable is introduced, as well as the measurement technique.
The various sources of systematic uncertainties are discussed in Sec.~\ref{sec:systematics}.
The results for the $\Lambda$ and $\overline\Lambda$ global polarization at two collision energies as a function of the hyperon transverse momentum and collision centrality are presented in Sec.~\ref{sec:results}.

\section{Data analysis}
\label{observable}

\subsection{The observable}

In this measurement, $\Lambda$ and $\overline\Lambda$ hyperons are reconstructed through their weak decay topologies \mbox{$\Lambda\rightarrow {\rm p}+\pi^-$} and $\overline\Lambda\rightarrow \bar {\rm p}+\pi^+$ (64\% branching ratio). The global polarization of the hyperons is determined from the angular distributions of their decay (anti-)protons.
In the hyperon rest frame, the (anti-)proton angular probability distribution, ${\rm d}w/{\rm d} \bm{n}^*_{\mathrm p}$, is given by 
\begin{equation}
\label{Eq:dWdn}
\frac{{\rm d}w}{{\rm d} \bm{n}^*_{\mathrm p}}= 1+\alpha_{\rm H}\,\bm{\zeta}_{\rm H}\cdot \bm{n}^*_{\mathrm p},
\end{equation}
where $\bm{n}^*_{\mathrm p}$ is the unit vector of the (anti-)proton direction.
A value from~\cite{Zyla:2020zbs} for the $\Lambda$ ($\bar\Lambda$)  decay parameter $\alpha_{\Lambda} = 0.732 \pm 0.014$ ($\alpha_{\bar\Lambda} = -0.758 \pm 0.012$) is used.
%Recent measurement by the BESIII Collaboration~\cite{Ablikim:2018zay} extracted a new values for $\alpha_{\Lambda}=0.750 \pm 0.010$ and $\alpha_{\bar\Lambda}=-0.758 \pm 0.012$, which are about 17\% larger than in~\cite{Tanabashi:2018oca}.
%Given the statistical and systematical uncertainties of the results reported below for the hyperon global polarization the new values for $\alpha_{\mathrm H}$ will not change the conclusions of this paper but they should be considered for future hight precision measurements.
% The $\alpha_{\mathrm H} = \pm\left(0.642 \pm 0.013\right)$, positive for $\Lambda$ and negative for $\overline\Lambda$, is the hyperon decay parameter~\cite{Tanabashi:2018oca}.
% The same absolute value of $\alpha_{\mathrm H}$ is used for $\Lambda$ and $\overline\Lambda$.
The polarization vector $\bm{\zeta}_{\mathrm H}$ in Eq.~(\ref{Eq:dWdn}), which is satisfying condition $|\bm{\zeta}_{\mathrm H}| \le 1$, can be measured experimentally as~\cite{Abelev:2007zk}
\begin{equation}
\label{eq2}
\bm{\zeta}_{\rm H} = \frac 3{\alpha_{\rm H}}\left<\bm{n}^*_{\rm p}\right>,
\end{equation}
where the brackets $\left<\dots\right>$ denote an event-by-event averaging over all hyperon decays.

The polarization vector $\bm{\zeta}_{\rm H}$ generally depends on the hyperon kinematics, namely the transverse momentum \pT, rapidity $y$, and its azimuthal angle with respect to the reaction plane, $\varphi-\PsiRP$, as well as the collision centrality.
The global polarization reported in this paper is determined by the component of the polarization vector perpendicular to the reaction plane.
The magnitude $\ph$ of the global polarization can be measured by averaging a corresponding projection of the $\bm{n}^*_{\rm p}$ vector, which in the laboratory coordinate system is given by $n^*_{{\rm p} \perp\rm RP}=\,\sin\theta^*_{\rm p}\,\sin\left(\varphi^*_{\rm p}-\PsiRP\right)$.
Here $\theta^*_{\rm p}$ ($\varphi^*_{\rm p}$) is the polar angle with respect to the collision axis (azimuthal angle) of the (anti-)proton direction in the hyperon rest frame.
Substituting $n^*_{p \perp\rm RP}$ into Eq.~(\ref{eq2}) and assuming an ideal detector acceptance, an average over the $\theta^*_{\rm p}$ yields
\begin{equation}
\label{eq3}
\ph = - \frac 8{\pi\alpha_{\rm H}}\left<\sin(\varphi^*_{\rm p}-\PsiRP)\right>.
\end{equation}
Here the brackets $\left<\dots\right>$ imply averaging over individual hyperons in all events.
The polarization is defined to be positive if the hyperon spin has a positive component along the system's angular momentum -- the same convention as employed in~\cite{STAR:2017ckg}.
The detector acceptance effects are treated in this work as systematic uncertainty and discussed in Sec.~\ref{sec:systematics}.

%-----------------------------------------------------------
%\subsection{Feed-down}

A significant fraction of $\Lambda$ and $\overline\Lambda$ hyperons originates from decay of heavier particles.
Existing estimates~\cite{Karpenko:2016jyx,Li:2017slc,Becattini:2016gvu} of the feed-down effect on the hyperon global polarization measurements, based on the assumption of thermal equilibrium and the particle yields from the statistical model~\cite{Wheaton:2004qb}, suggest that the primary $\Lambda$ and $\overline\Lambda$ polarization should be by about 15--20\% larger than what is measured at RHIC.
Taking into account that feed-down effects are expected to be small compared to other uncertainties in the current analysis, no correction to the data have been applied.

%--------------------------------------------------------------------
\subsection{Measurement technique}

The main components of the ALICE detector system~\cite{Aamodt:2008zz,Abelev:2014ffa} used for this measurement are the Time Projection Chamber (TPC) \cite{Dellacasa:2000bm}, the silicon detectors of the Inner Tracking System (ITS)~\cite{Dellacasa:1999kf} and the two neutron Zero-Degree Calorimeters (ZDC)~\cite{Dellacasa:1999ke}.
The analyzed data samples were recorded by ALICE in 2010 and 2011 for Pb--Pb collisions at $\sNN=2.76$~TeV, and in 2015 for collisions at $\sNN=5.02$~TeV.

During the data taking, the trigger required a hit in a pair of V0 detectors~\cite{Abbas:2013taa}.
%(scintillator arrays covering positive and negative pseudo-rapidity regions respectively 
In 2010 and 2011, events with only one V0 hit and at least two hits in the outer layer of the Silicon Pixel Detector (SPD) were accepted as well.
Contamination from beam-induced background was removed offline, as discussed in~\cite{Aamodt:2010pb,Aamodt:2010cz}.
The events with poor correlation between multiplicities in V0, ITS and TPC detectors were rejected.
The analysis was restricted to the events with the primary vertex along the beam direction, $V_z$, within $\pm10$~cm from the nominal center of the TPC.
This yielded for all collision centralities approximately 49 (75) million of Pb--Pb collisions at $\sNN=2.76$ (5.02)~TeV.
The collision centrality was determined using the energy deposition in the V0 detectors~\cite{Abelev:2013qoq}.

The reaction plane angle \PsiRP was estimated using the spectator plane angle $\psisp$, characterizing the deflection direction of the spectator neutrons.
The spectator deflections at positive and negative rapidity were reconstructed with a pair of neutron ZDC detectors located 114~m away from the interaction point.
The $\psisp$ angle was evaluated separately for the two detectors using the transverse profile of the spectator energy distribution provided by the $2\times 2$ segmentation of the ZDCs.
For this, a pair of two-dimensional vectors $\Qtp$ were constructed for two ZDCs following the procedure described in~\cite{Abelev:2013cva}:
\begin{equation}
\Qtp \equiv \left(\Xpm,\Ypm\right) =  \left.\sum\limits_{i=1}^{4}{\bf n}_i E_i^\tpspec\right/{\sum\limits_{i=1}^{4}} E_{i}^\tpspec,
\label{Eq:Qvector}
\end{equation}
where indices ${\rm p}$ and ${\rm t}$ denote the ZDC on the projectile ($\eta >0$) and target ($\eta<0$) side of the interaction point, respectively;
$E_i$ is the measured signal and ${\bf n}_i=\left(x_i,y_i\right)$ are the coordinates of the $i$-th ZDC segment.
The event averaged $\left<\Qtp\right>$ revealed a strong dependence on the collision centrality as well as on the three collision vertex coordinates, which is imposed by the offset of the LHC beam transverse spot positions relative to the nominal center of the ZDCs.
Moreover, the $\left<\Qtp\right>$ demonstrated a strong time dependence in some of the 2015 runs.
To compensate for these variations an event-by-event re-centering correction~\cite{Selyuzhenkov:2007zi} was applied as a function of collision centrality, three components of the collision vertex position, and beam time variation
\begin{equation}
\label{eq:QvectorRec}
\Qv' = \Qv-\mean{\Qv}.
\end{equation}

Additionally to the procedure described in~\cite{Abelev:2013cva}, the width of the $\Qv'$ distributions was equalized as a function of centrality, which together with Eq.~(\ref{eq:QvectorRec}) resulted in an overall $\Qv$-vector correction
\begin{equation}
\label{eq:QvectorCorr}
  \Qx'' = \frac{\Qx' - \left<\Qx\right>}{\left<\Qx'^2\right>},
  \quad
  \Qy'' = \frac{\Qy' - \left<\Qy\right>}{\left<\Qy'^2\right>}.
\end{equation}
The width equalization, up to 20\%, turned out to be particularly useful for the 2015 data sample, where one of the two neutron ZDC detectors lost signal from one of its four channels.

The $\psisp$ angle estimated from each ZDC is then given by the direction of the corrected $\Qv''$ vector
\begin{equation}
\label{eq:SPplane}
\Qv'' = |\Qv''| \exp\left\{i\psisp\right\}.
\end{equation}
To account for a finite resolution of the spectator plane angle $\psisp$, a correction $\Rep$ is introduce in the Eq.~(\ref{eq3}) following the method described in~\cite{Abelev:2007zk}
\begin{equation}
\label{eq:P_H}
\ph = - \frac 8{\pi\alpha_{\rm H}}\frac{\left<\sin(\varphi^*_{\rm p}-\psisp)\right>}{\Rep}.
\end{equation}
The polarization values obtained with Eq.~(\ref{eq:P_H}) for each of the two ZDC detectors were found to be similar within the statistical uncertainties and were combined.

\begin{figure}[ht]
\centering\includegraphics[width=0.65\hsize]{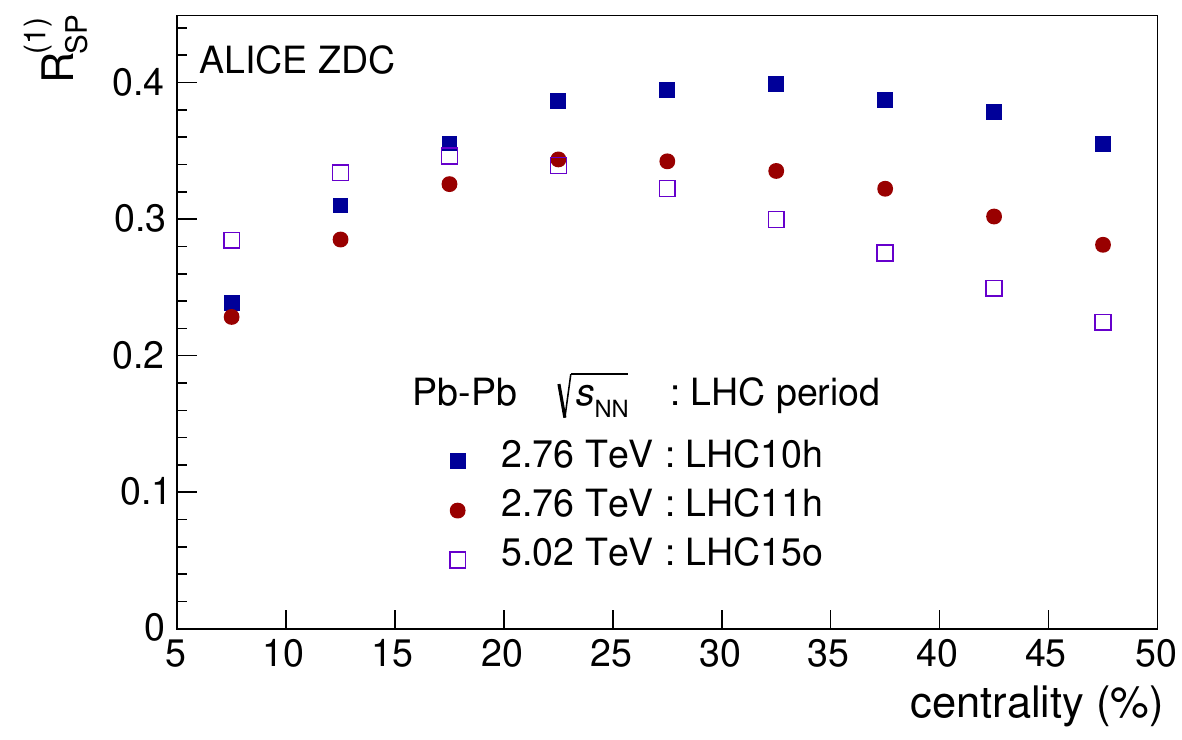}
\caption{
(color online)
The correction $\Rep$ for finite resolution of the spectator plane angle $\psisp$ as a function of collision centrality for three data sets used in the analysis.
Only statistical uncertainties, which are smaller than a symbol size, are shown.
}
\label{Fig:EPresolution}
\end{figure}

The correction \Rep was extracted from correlations between $\Qv$-vector angles from different ALICE detectors following the technique described in~\cite{Poskanzer:1998yz}.
Figure~\ref{Fig:EPresolution} presents the \Rep as a function of collision centrality for different data sets.
The most central (0--5\%) collisions are excluded from the analysis, because the small number of spectators does not allow for reliable estimation of the spectator deflection with the ZDCs.
During the Pb--Pb data taking in 2011 ($\sNN=2.76$~TeV) and 2015 ($\sNN=5.02$~TeV), the beams were collided at a non-zero vertical crossing angle~\cite{Abelev:2014ffa}.
This resulted in a partial screening of the spectator neutrons by the LHC tertiary collimators~\cite{Macina:2011} and in a degradation of the spectator plane resolution.
The V0s, TPC, and two forward multiplicity detectors (FMD)~\cite{FMD_TDR} were used to estimate a possible uncertainties in \Rep extraction for the two ZDC detectors.
In all data samples, these uncertainties turned out to be at a level of several percent.

The $\Lambda$ and $\overline\Lambda$ hyperons were reconstructed via their weak decay topology following the method and selection criteria described in~\cite{Abelev:2014ffa,Acharya:2018zuq}.
Charged daughter tracks from hyperon decay were required to have the pseudo-rapidity $|\eta| <0.9$ and at least 70 space points in the TPC.
The pion and (anti-)proton particle type assignments  were based on the track charge and specific energy loss (${\rm d}E/{\rm d}x$) measured in the TPC.
The daughter tracks were paired to form $\Lambda$ and $\overline\Lambda$ candidates.
The candidates were required to have transverse momentum \mbox{$\pt > 0.5$}~GeV/$c$, rapidity \mbox{$|y|<0.5$}, and the momentum vector pointing back to the primary collision vertex within a cone of opening angle less than~0.1 (0.08) radians for Pb--Pb collisions at $\sNN=2.76$ (5.02)~TeV.

The hyperon global polarization \ph was extracted with the fit to the measured correlation $\left<\sin(\varphi^*_{\rm p}-\psisp)\right>$ as a function of the $\Lambda$ or $\overline\Lambda$ candidate invariant mass, \Minv:
\begin{equation}
\label{eq:SinFit}
\left<\sin(\varphi^*_{\rm p}-\psisp)\right>(\Minv) = [1 - \fBG]\times S_\mathrm{H} +\fBG\times L_\mathrm{BG}(\Minv).
\end{equation}
Here the constant $S_\mathrm{H}$ gives the numerator of Eq.~(\ref{eq:P_H}) and $L(\Minv)$ is a linear parameterization of the background correlation as a function of \Minv.
The background fraction \fBG was evaluated as
\begin{equation}
\label{Eq:fBG}
 \fBG=\frac{N_{\rm BG}(\Minv)}{N_{\rm tot}(\Minv)}.
\end{equation}
Here $N_{\rm tot}(\Minv)$ is the total measured yield of the $\Lambda$ or $\overline\Lambda$ candidates.
The $N_{\rm BG}(\Minv)$ is given by the 4-th order polynomial fit to the $N_{\rm tot}(\Minv)$ outside of the $\Lambda$ and $\overline\Lambda$ invariant mass peak range, $\Minv < 1.107$~GeV/$c^2$ and $\Minv>1.125$~GeV/$c^2$.

\begin{figure}[ht]
\centering
\includegraphics[width=0.99\textwidth]{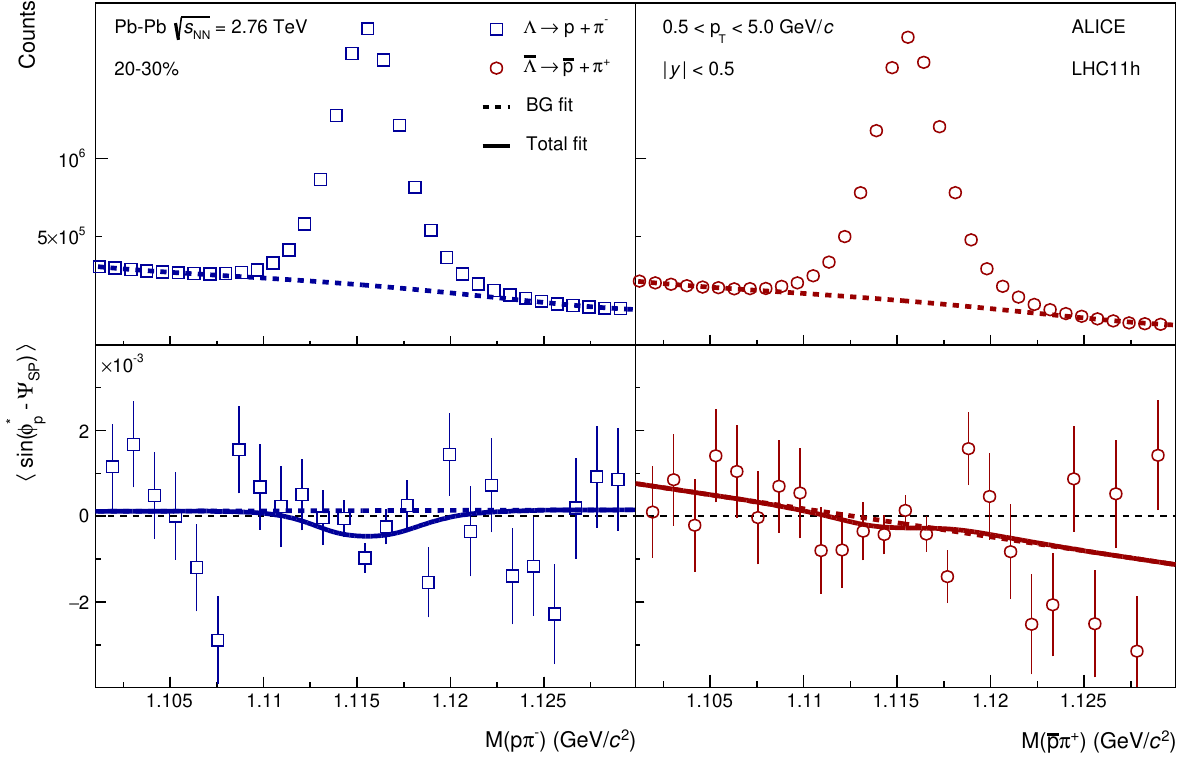}
\caption{
(color online)
(Upper panels) invariant mass distributions of $\Lambda$ (left) and $\overline\Lambda$ (right) candidates for 20--30\% centrality range in Pb--Pb collisions at $\sNN=2.76$~TeV using data collected during the LHC operation in 2011.
Dashed lines show the result of the fit for $N_{\rm BG}(\Minv)$ in Eq.~(\ref{Eq:fBG}).
(Bottom panels) Global polarization extraction via fit to $\left<\sin(\varphi^*_{\rm p}-\psisp)\right>$ as a function of the invariant mass, with $\psisp$ reconstructed with the ZDC on the target ($\eta<0$) side.
Dashed and solid lines show $L_\mathrm{BG}(\Minv)$ and the combined fit with Eq.~(\ref{eq:SinFit}).
See text for more details.
}
\label{Fig:InvMassMethod}
\end{figure}

The fitting procedure is illustrated in Fig.~\ref{Fig:InvMassMethod} for 20--30\% centrality range in Pb--Pb collisions at $\sNN=2.76$~TeV using data collected during the LHC operation in 2011.

\section{Systematic uncertainties}
\label{sec:systematics}

The considered sources of systematic uncertainties are summarized in Table~\ref{Tab:MainSystematics}.
The significance of a given systematical variation is determined following the procedure presented in~\cite{Barlow:2002yb}.
\begin{table}[ht]
\centering
\caption{Systematic uncertainties in global polarization measurements for different data sets. Values reported as a fraction of the corresponding statistical uncertainty.}
\begin{tabular}{|c|c|c|l|c|l|}
\hline
Data sample & centrality & \makecell{$\Lambda$/$\overline\Lambda$ candidate\\selection} & \makecell{Fitting\\procedure} & Acceptance & $\Rep$\\
\hline
\hline
2010 & \makecell{5--15\%\\ 15--50\%} & 20\% & \makecell{30\% \\ ~~5\%} & 10\% & $<2\%$\\
\hline
2011 & \makecell{5--15\%\\ 15--50\%} & 30\% & \makecell{20\%  \\ 12\% } & 10\% & $<2$\%\\
\hline
2015 & 5--50\% & 30\% & ~~~~~20\% & 10\% & $<4$\% \\
\hline
\end{tabular}
\label{Tab:MainSystematics}
\end{table}

The selection on the primary vertex position $V_z$ was varied to $\pm7$~cm and $\pm8$~cm instead of the default $\pm10$~cm.
%Each of the pile-up rejection criteria were tightened by about 10\% of its default value.
The analysis was repeated with the collision centrality estimated with either ITS or TPC detectors (instead of default V0 detector).
Variations of the results when changing the event selection criteria described above and centrality estimators are found to be negligible and were not included into the total systematic uncertainty.

The following $\Lambda$ and $\overline\Lambda$ candidate selection criteria were varied:
the distance of closest approach to the primary vertex, decay daughter track selection via specific energy loss in the TPC, the distance of closest approach between the pair of the daughter tracks, and the criteria on the hyperon candidates momentum pointing angle to the primary vertex.
The corresponding contribution from these variations to the total systematic uncertainty is about 20--30\% of the statistical uncertainty.
The main contribution to the systematic uncertainty from the fitting procedure in Eq.~(\ref{eq:SinFit}) comes from a variation of the fit region.

In Eq.~(\ref{eq:P_H}) a perfect acceptance of the hyperons and their decay daughters is assumed.
In the case of an imperfect detector, there is a overall scale correction of the measured polarization as well as a possible admixture of higher-order harmonics in a Fourier decomposition of the $\mathrm{d}N_{\Lambda/\overline{\Lambda}}/\mathrm{d}(\varphi^*_{\rm p}-\PsiRP)$ distribution.
Using the method described in~\cite{Selyuzhenkov:2006tj,Abelev:2007zk} the corresponding relative uncertainty was estimated to be about 10\% independent of centrality.
This uncertainty comes primarily from the admixture of the higher-order harmonics when hyperon $\pT\lesssim 2$~GeV/$c$.

A detailed study was performed for the evaluation of the systematic uncertainties of $\Rep$ as well as for the evaluation of the difference between the two neutron ZDC detectors.
The correlations between the flow vectors from different detectors, including TPC and pairs of ZDC, V0, FMD detectors were studied.
The contribution the total systematic uncertainties due to the $\Rep$ extraction were found to be at a level of a few percent.

\section{Results}
\label{sec:results}

\begin{figure}[ht]
\centering\includegraphics[width=0.99\textwidth]{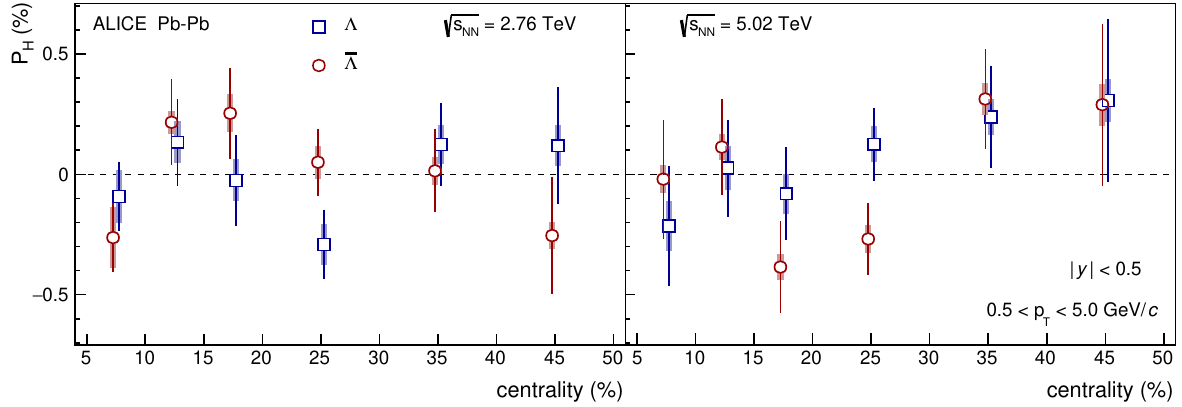}
\caption{
(color online)
The global hyperon polarization as function of centrality for Pb--Pb collisions at $\sNN=2.76$~TeV (left) and $5.02$~TeV (right).
The systematic uncertainties are shown as shaded boxes.
Points are slightly shifted along the horizontal axis for better visibility.
}
\label{Fig:ResultsCentrality}
\end{figure}

\begin{figure}[ht]
\centering\includegraphics[width=0.99\textwidth]{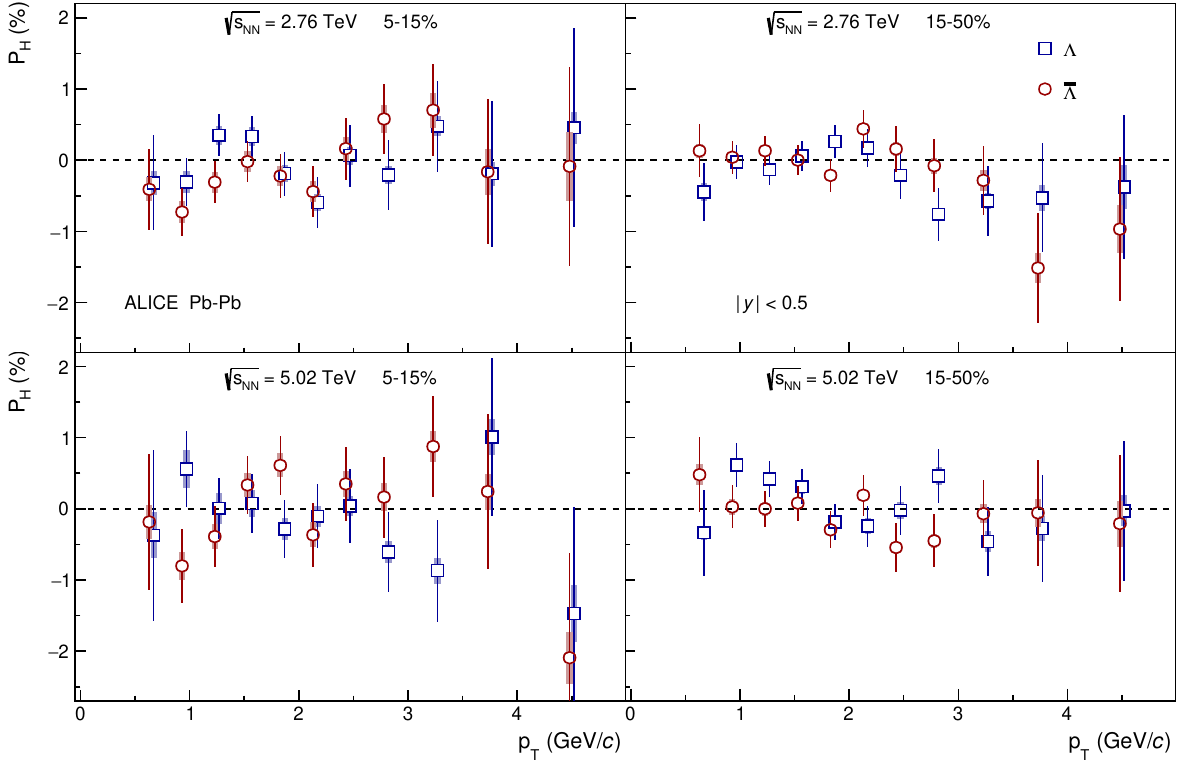}
\caption{
(color online)
The global hyperon polarization as function of transverse momentum \pT for Pb--Pb collisions at $\sNN=2.76$~TeV (upper) and $5.02$~TeV (lower) in 5--15\% (left) and 15--50\% (right) centrality classes.
The systematic uncertainties are shown as shaded boxes.
Points are slightly shifted along the horizontal axis for better visibility.
}
\label{Fig:ResultsPt}
\end{figure}

Figures~\ref{Fig:ResultsCentrality} and \ref{Fig:ResultsPt} show the measured hyperon global polarization \ph as a function of centrality and hyperon transverse momentum, \pT, in Pb--Pb collisions for two collision energies.
The results from 2010 and 2011 data samples were combined accounting for the corresponding statistical and systematic uncertainty.
At RHIC energies, the global polarization exhibited a clear centrality dependence with three times large magnitude in peripheral collisions compared to that in central, while no significant \pT dependence within the accessible \pT range was observed~\cite{Adam:2018ivw}.
The \ph at the LHC is found to be consistent with zero  within the experimental uncertainties for all studied centrality classes (Fig.~\ref{Fig:ResultsCentrality}) and \pT ranges (Fig.~\ref{Fig:ResultsPt}).
By repeating a similar analysis, no signal signal was observed as a function of rapidity either.

\begin{figure}[ht]
\center
\includegraphics[width=0.75\textwidth]{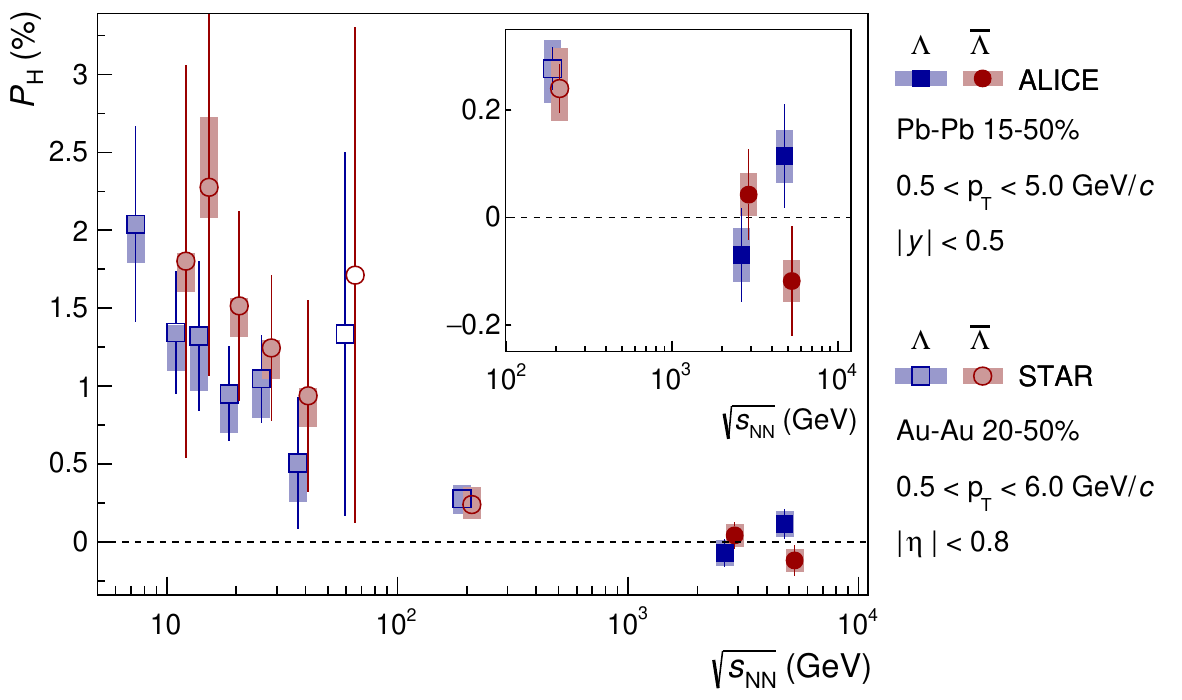}
\caption{
(color online)
The global hyperon polarization as a function of collision energy.
Results are compared with the STAR data at lower energies~\cite{STAR:2017ckg,Adam:2018ivw}.
The insert shows zoomed-in comparison with the data at the top RHIC energy.
The systematic uncertainties are shown as shaded boxes.
Points are slightly shifted along the horizontal axis for better visibility.
}
\label{fig:final}
\end{figure}

The average global polarization for two centrality ranges, 5--15\% and 15--50\%, are presented in Fig.~\ref{fig:final}, while numerical values are reported in Tab.~\ref{tab:average}.
Figure~\ref{fig:final} also presents the comparison with the STAR data~\cite{STAR:2017ckg,Adam:2018ivw} for lower collision energies.
Despite large uncertainties, the ALICE measurements confirm the trend of the global polarization decreasing with increasing collision energy.

Assuming the same values of the global polarization for $\Lambda$ and $\overline\Lambda$ and neglecting the possible difference of about 30\% (according to the empirical estimates discussed above) between the two LHC energies, one can average all four ALICE data points for 15--50\% centrality, where the largest signal is expected.
This yields a value $\mean{\ph}(\%)\approx - 0.01\pm 0.05~\mbox{(stat.)} \pm 0.03~\mbox{(syst.)}$ for 15--50\% centrality, which is consistent with the empirical estimates of $\ph(\%) \approx 0.04-0.08$ based on directed flow measurements.
%
%----------------------------- Table average
\begin{table}[ht]
\caption{The global polarization of $\Lambda$ and $\overline\Lambda$ hyperons in Pb--Pb collisions at $\sNN=2.76$~TeV and $5.02$~TeV for centrality ranges 5--15\% and 15--50\%.}
\centering
\begin{tabular}{|c|c|c|c|}
\hline
  $\sNN$ & centrality& $P_\Lambda \,(\%)$ & $P_{\overline\Lambda} \,(\%)$ \\
 \hline
\hline
 \multirow{2}{*}{2.76 TeV} &
  5--15\% &
    \hskip 3mm $-0.01 \pm 0.12~\mbox{(stat.)}\pm 0.04~\mbox{(syst.)}$ &
   \hskip 3mm$-0.08 \pm 0.12~\mbox{(stat.)} \pm 0.07~\mbox{(syst.)}$ \\
 & 15--50\% &
       \hskip 3mm$-0.07 \pm 0.09~\mbox{(stat.)}\pm 0.04~\mbox{(syst.)}$ &
            $0.05 \pm 0.09~\mbox{(stat.)}\pm 0.03~\mbox{(syst.)}$ \\
  \hline
 \multirow{2}{*}{5.02 TeV} &

 5--15\% &
     \hskip  3mm$-0.07 \pm 0.16~\mbox{(stat.)} \pm 0.07~\mbox{(syst.)}$ &
       $0.06 \pm 0.16~\mbox{(stat.)} \pm 0.03~\mbox{(syst.)}$ \\
 & 15--50\%&
      $0.12 \pm 0.10~\mbox{(stat.)}\pm 0.04~\mbox{(syst.)}$ &
       \hskip 3mm     $-0.13 \pm 0.11~\mbox{(stat.)}\pm 0.03~\mbox{(syst.)}$\\
  \hline
 average &
  15--50\%&
 \multicolumn{2}{c|}{$\mean{\ph}(\%)\approx -0.01\pm 0.05~\mbox{(stat.)} \pm 0.03~\mbox{(syst.)}$}\\
  \hline
\end{tabular}
\label{tab:average}
\end{table}

\section{Summary}

The first measurements of $\Lambda$ and $\overline\Lambda$ hyperons global polarization are reported for Pb--Pb collisions at $\sNN=2.76$ and 5.02~TeV recorded with ALICE at the LHC.
The hyperon global polarization has been measured differentially as a function of centrality and transverse momentum (\pT) for the range of collision centrality 5--50\%, $0.5 < \pT<5$~GeV/$c$, and rapidity $|y|<0.5$.
No significant dependences neither splitting between the global polarization values for $\Lambda$ and $\overline\Lambda$ has been observed.
The average $\Lambda$ and $\overline\Lambda$ polarization for 15--50\% centrality range at two collision energies is found to be consistent with zero, $\mean{\ph}(\%)\approx -0.01\pm 0.05~\mbox{(stat.)}  \pm 0.03~\mbox{(syst.)}$.
This confirms the observed earlier trend of the global polarization decrease with increasing collision energy.
The results are compatible with the extrapolation of the RHIC results and empirical estimates of $\ph(\%) \approx 0.04-0.08$ based on the similarity of collision energy dependence of the global polarization and the slope of the directed flow in the midrapidity region.
The high luminosity LHC run after the 2019-2021 long shutdown with the upgraded ALICE detector will bring more than 100 times more data, which should allow tests of the previously mentioned prediction with much better accuracy.

%\input{Erratum}

%===========================================================

%%%%% acknowledgements - handled by EB chairs 
\newenvironment{acknowledgement}{\relax}{\relax}
\begin{acknowledgement}
\section*{Acknowledgements}
% add specific acknowledgements here 
% ...but please don't remove the line below: funding agencies
% will be acknowledged with a custom tex file handled by EB chairs after Collab Round 2
% Version: 2019-07-15

The ALICE Collaboration would like to thank all its engineers and technicians for their invaluable contributions to the construction of the experiment and the CERN accelerator teams for the outstanding performance of the LHC complex.
The ALICE Collaboration gratefully acknowledges the resources and support provided by all Grid centres and the Worldwide LHC Computing Grid (WLCG) collaboration.
The ALICE Collaboration acknowledges the following funding agencies for their support in building and running the ALICE detector:
A. I. Alikhanyan National Science Laboratory (Yerevan Physics Institute) Foundation (ANSL), State Committee of Science and World Federation of Scientists (WFS), Armenia;
Austrian Academy of Sciences, Austrian Science Fund (FWF): [M 2467-N36] and Nationalstiftung f\"{u}r Forschung, Technologie und Entwicklung, Austria;
Ministry of Communications and High Technologies, National Nuclear Research Center, Azerbaijan;
Conselho Nacional de Desenvolvimento Cient\'{\i}fico e Tecnol\'{o}gico (CNPq), Financiadora de Estudos e Projetos (Finep), Funda\c{c}\~{a}o de Amparo \`{a} Pesquisa do Estado de S\~{a}o Paulo (FAPESP) and Universidade Federal do Rio Grande do Sul (UFRGS), Brazil;
Ministry of Education of China (MOEC) , Ministry of Science \& Technology of China (MSTC) and National Natural Science Foundation of China (NSFC), China;
Ministry of Science and Education and Croatian Science Foundation, Croatia;
Centro de Aplicaciones Tecnol\'{o}gicas y Desarrollo Nuclear (CEADEN), Cubaenerg\'{\i}a, Cuba;
Ministry of Education, Youth and Sports of the Czech Republic, Czech Republic;
The Danish Council for Independent Research | Natural Sciences, the VILLUM FONDEN and Danish National Research Foundation (DNRF), Denmark;
Helsinki Institute of Physics (HIP), Finland;
Commissariat \`{a} l'Energie Atomique (CEA), Institut National de Physique Nucl\'{e}aire et de Physique des Particules (IN2P3) and Centre National de la Recherche Scientifique (CNRS) and R\'{e}gion des  Pays de la Loire, France;
Bundesministerium f\"{u}r Bildung und Forschung (BMBF) and GSI Helmholtzzentrum f\"{u}r Schwerionenforschung GmbH, Germany;
General Secretariat for Research and Technology, Ministry of Education, Research and Religions, Greece;
National Research, Development and Innovation Office, Hungary;
Department of Atomic Energy Government of India (DAE), Department of Science and Technology, Government of India (DST), University Grants Commission, Government of India (UGC) and Council of Scientific and Industrial Research (CSIR), India;
Indonesian Institute of Science, Indonesia;
Centro Fermi - Museo Storico della Fisica e Centro Studi e Ricerche Enrico Fermi and Istituto Nazionale di Fisica Nucleare (INFN), Italy;
Institute for Innovative Science and Technology , Nagasaki Institute of Applied Science (IIST), Japanese Ministry of Education, Culture, Sports, Science and Technology (MEXT) and Japan Society for the Promotion of Science (JSPS) KAKENHI, Japan;
Consejo Nacional de Ciencia (CONACYT) y Tecnolog\'{i}a, through Fondo de Cooperaci\'{o}n Internacional en Ciencia y Tecnolog\'{i}a (FONCICYT) and Direcci\'{o}n General de Asuntos del Personal Academico (DGAPA), Mexico;
Nederlandse Organisatie voor Wetenschappelijk Onderzoek (NWO), Netherlands;
The Research Council of Norway, Norway;
Commission on Science and Technology for Sustainable Development in the South (COMSATS), Pakistan;
Pontificia Universidad Cat\'{o}lica del Per\'{u}, Peru;
Ministry of Science and Higher Education and National Science Centre, Poland;
Korea Institute of Science and Technology Information and National Research Foundation of Korea (NRF), Republic of Korea;
Ministry of Education and Scientific Research, Institute of Atomic Physics and Ministry of Research and Innovation and Institute of Atomic Physics, Romania;
Joint Institute for Nuclear Research (JINR), Ministry of Education and Science of the Russian Federation, National Research Centre Kurchatov Institute, Russian Science Foundation and Russian Foundation for Basic Research, Russia;
Ministry of Education, Science, Research and Sport of the Slovak Republic, Slovakia;
National Research Foundation of South Africa, South Africa;
Swedish Research Council (VR) and Knut \& Alice Wallenberg Foundation (KAW), Sweden;
European Organization for Nuclear Research, Switzerland;
Suranaree University of Technology (SUT), National Science and Technology Development Agency (NSDTA) and Office of the Higher Education Commission under NRU project of Thailand, Thailand;
Turkish Atomic Energy Agency (TAEK), Turkey;
National Academy of  Sciences of Ukraine, Ukraine;
Science and Technology Facilities Council (STFC), United Kingdom;
National Science Foundation of the United States of America (NSF) and United States Department of Energy, Office of Nuclear Physics (DOE NP), United States of America.
\end{acknowledgement}

%%%%%%%% Bibliography 
\bibliographystyle{utphys}   % Remember we use title in the biblio
\bibliography{References}
%\input {bibliography.tex}  

%%%%%%%%%%%%%%%%%%%%%%%%%%%%%%%%
% Appendices: yours (if any) + authorlist
%%%%%%%%%%%%%%%%%%%%%%%%%%%%%%%%
\newpage
\appendix

%
%\input{} % put your appendices here (if any)
%

%%%%% Authorlist - please do not touch: handled by EB chairs 
\section{The ALICE Collaboration}
\label{app:collab}
% Collaboration: CERN-LHC-ALICE
% Generation Date is 2019-Jul-15

% How to use:
%%%%%%%%% appendix with author list
%\appendix
%\section{The ALICE Collaboration}
%\label{app:collab}
%\input{Alice_Authorslist_XXXX-Axx-XX.tex}
\begingroup
\small
\begin{flushleft}
S.~Acharya\Irefn{org141}\And 
D.~Adamov\'{a}\Irefn{org93}\And 
S.P.~Adhya\Irefn{org141}\And 
A.~Adler\Irefn{org73}\And 
J.~Adolfsson\Irefn{org79}\And 
M.M.~Aggarwal\Irefn{org98}\And 
G.~Aglieri Rinella\Irefn{org34}\And 
M.~Agnello\Irefn{org31}\And 
N.~Agrawal\Irefn{org10}\textsuperscript{,}\Irefn{org48}\textsuperscript{,}\Irefn{org53}\And 
Z.~Ahammed\Irefn{org141}
% ...
\And
\renewcommand\labelenumi{\textsuperscript{\theenumi}~}

\section*{Affiliation notes}
\renewcommand\theenumi{\roman{enumi}}
\begin{Authlist}
\item \Adef{org*}Deceased
\item \Adef{orgI}Dipartimento DET del Politecnico di Torino, Turin, Italy
\item \Adef{orgII}M.V. Lomonosov Moscow State University, D.V. Skobeltsyn Institute of Nuclear, Physics, Moscow, Russia
\item \Adef{orgIII}Department of Applied Physics, Aligarh Muslim University, Aligarh, India
\item \Adef{orgIV}Institute of Theoretical Physics, University of Wroclaw, Poland
\end{Authlist}

\section*{Collaboration Institutes}
\renewcommand\theenumi{\arabic{enumi}~}
\begin{Authlist}
\item \Idef{org1}A.I. Alikhanyan National Science Laboratory (Yerevan Physics Institute) Foundation, Yerevan, Armenia
\item \Idef{org2}Bogolyubov Institute for Theoretical Physics, National Academy of Sciences of Ukraine, Kiev, Ukraine
\item \Idef{org3}Bose Institute, Department of Physics  and Centre for Astroparticle Physics and Space Science (CAPSS), Kolkata, India
\item \Idef{org4}Budker Institute for Nuclear Physics, Novosibirsk, Russia
\item \Idef{org5}California Polytechnic State University, San Luis Obispo, California, United States
\item \Idef{org6}Central China Normal University, Wuhan, China
\item \Idef{org7}Centre de Calcul de l'IN2P3, Villeurbanne, Lyon, France
\item \Idef{org8}Centro de Aplicaciones Tecnol\'{o}gicas y Desarrollo Nuclear (CEADEN), Havana, Cuba
\item \Idef{org9}Centro de Investigaci\'{o}n y de Estudios Avanzados (CINVESTAV), Mexico City and M\'{e}rida, Mexico
\item \Idef{org10}Centro Fermi - Museo Storico della Fisica e Centro Studi e Ricerche ``Enrico Fermi', Rome, Italy
\item \Idef{org11}Chicago State University, Chicago, Illinois, United States
\item \Idef{org12}China Institute of Atomic Energy, Beijing, China
\item \Idef{org13}Chonbuk National University, Jeonju, Republic of Korea
\item \Idef{org14}Comenius University Bratislava, Faculty of Mathematics, Physics and Informatics, Bratislava, Slovakia
\item \Idef{org15}COMSATS University Islamabad, Islamabad, Pakistan
\item \Idef{org16}Creighton University, Omaha, Nebraska, United States
\item \Idef{org17}Department of Physics, Aligarh Muslim University, Aligarh, India
\item \Idef{org18}Department of Physics, Pusan National University, Pusan, Republic of Korea
\item \Idef{org19}Department of Physics, Sejong University, Seoul, Republic of Korea
\item \Idef{org20}Department of Physics, University of California, Berkeley, California, United States
\item \Idef{org21}Department of Physics, University of Oslo, Oslo, Norway
\item \Idef{org22}Department of Physics and Technology, University of Bergen, Bergen, Norway
\item \Idef{org23}Dipartimento di Fisica dell'Universit\`{a} 'La Sapienza' and Sezione INFN, Rome, Italy
\item \Idef{org24}Dipartimento di Fisica dell'Universit\`{a} and Sezione INFN, Cagliari, Italy
\item \Idef{org25}Dipartimento di Fisica dell'Universit\`{a} and Sezione INFN, Trieste, Italy
\item \Idef{org26}Dipartimento di Fisica dell'Universit\`{a} and Sezione INFN, Turin, Italy
\item \Idef{org27}Dipartimento di Fisica e Astronomia dell'Universit\`{a} and Sezione INFN, Bologna, Italy
\item \Idef{org28}Dipartimento di Fisica e Astronomia dell'Universit\`{a} and Sezione INFN, Catania, Italy
\item \Idef{org29}Dipartimento di Fisica e Astronomia dell'Universit\`{a} and Sezione INFN, Padova, Italy
\item \Idef{org30}Dipartimento di Fisica `E.R.~Caianiello' dell'Universit\`{a} and Gruppo Collegato INFN, Salerno, Italy
\item \Idef{org31}Dipartimento DISAT del Politecnico and Sezione INFN, Turin, Italy
\item \Idef{org32}Dipartimento di Scienze e Innovazione Tecnologica dell'Universit\`{a} del Piemonte Orientale and INFN Sezione di Torino, Alessandria, Italy
\item \Idef{org33}Dipartimento Interateneo di Fisica `M.~Merlin' and Sezione INFN, Bari, Italy
\item \Idef{org34}European Organization for Nuclear Research (CERN), Geneva, Switzerland
\item \Idef{org35}Faculty of Electrical Engineering, Mechanical Engineering and Naval Architecture, University of Split, Split, Croatia
\item \Idef{org36}Faculty of Engineering and Science, Western Norway University of Applied Sciences, Bergen, Norway
\item \Idef{org37}Faculty of Nuclear Sciences and Physical Engineering, Czech Technical University in Prague, Prague, Czech Republic
\item \Idef{org38}Faculty of Science, P.J.~\v{S}af\'{a}rik University, Ko\v{s}ice, Slovakia
\item \Idef{org39}Frankfurt Institute for Advanced Studies, Johann Wolfgang Goethe-Universit\"{a}t Frankfurt, Frankfurt, Germany
\item \Idef{org40}Gangneung-Wonju National University, Gangneung, Republic of Korea
\item \Idef{org41}Gauhati University, Department of Physics, Guwahati, India
\item \Idef{org42}Helmholtz-Institut f\"{u}r Strahlen- und Kernphysik, Rheinische Friedrich-Wilhelms-Universit\"{a}t Bonn, Bonn, Germany
\item \Idef{org43}Helsinki Institute of Physics (HIP), Helsinki, Finland
\item \Idef{org44}High Energy Physics Group,  Universidad Aut\'{o}noma de Puebla, Puebla, Mexico
\item \Idef{org45}Hiroshima University, Hiroshima, Japan
\item \Idef{org46}Hochschule Worms, Zentrum  f\"{u}r Technologietransfer und Telekommunikation (ZTT), Worms, Germany
\item \Idef{org47}Horia Hulubei National Institute of Physics and Nuclear Engineering, Bucharest, Romania
\item \Idef{org48}Indian Institute of Technology Bombay (IIT), Mumbai, India
\item \Idef{org49}Indian Institute of Technology Indore, Indore, India
\item \Idef{org50}Indonesian Institute of Sciences, Jakarta, Indonesia
\item \Idef{org51}INFN, Laboratori Nazionali di Frascati, Frascati, Italy
\item \Idef{org52}INFN, Sezione di Bari, Bari, Italy
\item \Idef{org53}INFN, Sezione di Bologna, Bologna, Italy
\item \Idef{org54}INFN, Sezione di Cagliari, Cagliari, Italy
\item \Idef{org55}INFN, Sezione di Catania, Catania, Italy
\item \Idef{org56}INFN, Sezione di Padova, Padova, Italy
\item \Idef{org57}INFN, Sezione di Roma, Rome, Italy
\item \Idef{org58}INFN, Sezione di Torino, Turin, Italy
\item \Idef{org59}INFN, Sezione di Trieste, Trieste, Italy
\item \Idef{org60}Inha University, Incheon, Republic of Korea
\item \Idef{org61}Institut de Physique Nucl\'{e}aire d'Orsay (IPNO), Institut National de Physique Nucl\'{e}aire et de Physique des Particules (IN2P3/CNRS), Universit\'{e} de Paris-Sud, Universit\'{e} Paris-Saclay, Orsay, France
\item \Idef{org62}Institute for Nuclear Research, Academy of Sciences, Moscow, Russia
\item \Idef{org63}Institute for Subatomic Physics, Utrecht University/Nikhef, Utrecht, Netherlands
\item \Idef{org64}Institute of Experimental Physics, Slovak Academy of Sciences, Ko\v{s}ice, Slovakia
\item \Idef{org65}Institute of Physics, Homi Bhabha National Institute, Bhubaneswar, India
\item \Idef{org66}Institute of Physics of the Czech Academy of Sciences, Prague, Czech Republic
\item \Idef{org67}Institute of Space Science (ISS), Bucharest, Romania
\item \Idef{org68}Institut f\"{u}r Kernphysik, Johann Wolfgang Goethe-Universit\"{a}t Frankfurt, Frankfurt, Germany
\item \Idef{org69}Instituto de Ciencias Nucleares, Universidad Nacional Aut\'{o}noma de M\'{e}xico, Mexico City, Mexico
\item \Idef{org70}Instituto de F\'{i}sica, Universidade Federal do Rio Grande do Sul (UFRGS), Porto Alegre, Brazil
\item \Idef{org71}Instituto de F\'{\i}sica, Universidad Nacional Aut\'{o}noma de M\'{e}xico, Mexico City, Mexico
\item \Idef{org72}iThemba LABS, National Research Foundation, Somerset West, South Africa
\item \Idef{org73}Johann-Wolfgang-Goethe Universit\"{a}t Frankfurt Institut f\"{u}r Informatik, Fachbereich Informatik und Mathematik, Frankfurt, Germany
\item \Idef{org74}Joint Institute for Nuclear Research (JINR), Dubna, Russia
\item \Idef{org75}Korea Institute of Science and Technology Information, Daejeon, Republic of Korea
\item \Idef{org76}KTO Karatay University, Konya, Turkey
\item \Idef{org77}Laboratoire de Physique Subatomique et de Cosmologie, Universit\'{e} Grenoble-Alpes, CNRS-IN2P3, Grenoble, France
\item \Idef{org78}Lawrence Berkeley National Laboratory, Berkeley, California, United States
\item \Idef{org79}Lund University Department of Physics, Division of Particle Physics, Lund, Sweden
\item \Idef{org80}Nagasaki Institute of Applied Science, Nagasaki, Japan
\item \Idef{org81}Nara Women{'}s University (NWU), Nara, Japan
\item \Idef{org82}National and Kapodistrian University of Athens, School of Science, Department of Physics , Athens, Greece
\item \Idef{org83}National Centre for Nuclear Research, Warsaw, Poland
\item \Idef{org84}National Institute of Science Education and Research, Homi Bhabha National Institute, Jatni, India
\item \Idef{org85}National Nuclear Research Center, Baku, Azerbaijan
\item \Idef{org86}National Research Centre Kurchatov Institute, Moscow, Russia
\item \Idef{org87}Niels Bohr Institute, University of Copenhagen, Copenhagen, Denmark
\item \Idef{org88}Nikhef, National institute for subatomic physics, Amsterdam, Netherlands
\item \Idef{org89}NRC Kurchatov Institute IHEP, Protvino, Russia
\item \Idef{org90}NRC \guillemotleft Kurchatov\guillemotright~Institute - ITEP, Moscow, Russia
\item \Idef{org91}NRNU Moscow Engineering Physics Institute, Moscow, Russia
\item \Idef{org92}Nuclear Physics Group, STFC Daresbury Laboratory, Daresbury, United Kingdom
\item \Idef{org93}Nuclear Physics Institute of the Czech Academy of Sciences, \v{R}e\v{z} u Prahy, Czech Republic
\item \Idef{org94}Oak Ridge National Laboratory, Oak Ridge, Tennessee, United States
\item \Idef{org95}Ohio State University, Columbus, Ohio, United States
\item \Idef{org96}Petersburg Nuclear Physics Institute, Gatchina, Russia
\item \Idef{org97}Physics department, Faculty of science, University of Zagreb, Zagreb, Croatia
\item \Idef{org98}Physics Department, Panjab University, Chandigarh, India
\item \Idef{org99}Physics Department, University of Jammu, Jammu, India
\item \Idef{org100}Physics Department, University of Rajasthan, Jaipur, India
\item \Idef{org101}Physikalisches Institut, Eberhard-Karls-Universit\"{a}t T\"{u}bingen, T\"{u}bingen, Germany
\item \Idef{org102}Physikalisches Institut, Ruprecht-Karls-Universit\"{a}t Heidelberg, Heidelberg, Germany
\item \Idef{org103}Physik Department, Technische Universit\"{a}t M\"{u}nchen, Munich, Germany
\item \Idef{org104}Politecnico di Bari, Bari, Italy
\item \Idef{org105}Research Division and ExtreMe Matter Institute EMMI, GSI Helmholtzzentrum f\"ur Schwerionenforschung GmbH, Darmstadt, Germany
\item \Idef{org106}Rudjer Bo\v{s}kovi\'{c} Institute, Zagreb, Croatia
\item \Idef{org107}Russian Federal Nuclear Center (VNIIEF), Sarov, Russia
\item \Idef{org108}Saha Institute of Nuclear Physics, Homi Bhabha National Institute, Kolkata, India
\item \Idef{org109}School of Physics and Astronomy, University of Birmingham, Birmingham, United Kingdom
\item \Idef{org110}Secci\'{o}n F\'{\i}sica, Departamento de Ciencias, Pontificia Universidad Cat\'{o}lica del Per\'{u}, Lima, Peru
\item \Idef{org111}Shanghai Institute of Applied Physics, Shanghai, China
\item \Idef{org112}St. Petersburg State University, St. Petersburg, Russia
\item \Idef{org113}Stefan Meyer Institut f\"{u}r Subatomare Physik (SMI), Vienna, Austria
\item \Idef{org114}SUBATECH, IMT Atlantique, Universit\'{e} de Nantes, CNRS-IN2P3, Nantes, France
\item \Idef{org115}Suranaree University of Technology, Nakhon Ratchasima, Thailand
\item \Idef{org116}Technical University of Ko\v{s}ice, Ko\v{s}ice, Slovakia
\item \Idef{org117}Technische Universit\"{a}t M\"{u}nchen, Excellence Cluster 'Universe', Munich, Germany
\item \Idef{org118}The Henryk Niewodniczanski Institute of Nuclear Physics, Polish Academy of Sciences, Cracow, Poland
\item \Idef{org119}The University of Texas at Austin, Austin, Texas, United States
\item \Idef{org120}Universidad Aut\'{o}noma de Sinaloa, Culiac\'{a}n, Mexico
\item \Idef{org121}Universidade de S\~{a}o Paulo (USP), S\~{a}o Paulo, Brazil
\item \Idef{org122}Universidade Estadual de Campinas (UNICAMP), Campinas, Brazil
\item \Idef{org123}Universidade Federal do ABC, Santo Andre, Brazil
\item \Idef{org124}University of Cape Town, Cape Town, South Africa
\item \Idef{org125}University of Houston, Houston, Texas, United States
\item \Idef{org126}University of Jyv\"{a}skyl\"{a}, Jyv\"{a}skyl\"{a}, Finland
\item \Idef{org127}University of Liverpool, Liverpool, United Kingdom
\item \Idef{org128}University of Science and Techonology of China, Hefei, China
\item \Idef{org129}University of South-Eastern Norway, Tonsberg, Norway
\item \Idef{org130}University of Tennessee, Knoxville, Tennessee, United States
\item \Idef{org131}University of the Witwatersrand, Johannesburg, South Africa
\item \Idef{org132}University of Tokyo, Tokyo, Japan
\item \Idef{org133}University of Tsukuba, Tsukuba, Japan
\item \Idef{org134}Universit\'{e} Clermont Auvergne, CNRS/IN2P3, LPC, Clermont-Ferrand, France
\item \Idef{org135}Universit\'{e} de Lyon, Universit\'{e} Lyon 1, CNRS/IN2P3, IPN-Lyon, Villeurbanne, Lyon, France
\item \Idef{org136}Universit\'{e} de Strasbourg, CNRS, IPHC UMR 7178, F-67000 Strasbourg, France, Strasbourg, France
\item \Idef{org137}Universit\'{e} Paris-Saclay Centre d'Etudes de Saclay (CEA), IRFU, D\'{e}partment de Physique Nucl\'{e}aire (DPhN), Saclay, France
\item \Idef{org138}Universit\`{a} degli Studi di Foggia, Foggia, Italy
\item \Idef{org139}Universit\`{a} degli Studi di Pavia, Pavia, Italy
\item \Idef{org140}Universit\`{a} di Brescia, Brescia, Italy
\item \Idef{org141}Variable Energy Cyclotron Centre, Homi Bhabha National Institute, Kolkata, India
\item \Idef{org142}Warsaw University of Technology, Warsaw, Poland
\item \Idef{org143}Wayne State University, Detroit, Michigan, United States
\item \Idef{org144}Westf\"{a}lische Wilhelms-Universit\"{a}t M\"{u}nster, Institut f\"{u}r Kernphysik, M\"{u}nster, Germany
\item \Idef{org145}Wigner Research Centre for Physics, Hungarian Academy of Sciences, Budapest, Hungary
\item \Idef{org146}Yale University, New Haven, Connecticut, United States
\item \Idef{org147}Yonsei University, Seoul, Republic of Korea
\end{Authlist}
\endgroup
  
\end{document}